\documentclass[10pt,twocolumn]{article}
\usepackage{todonotes}
\usepackage{authblk}
\usepackage{cite}
\usepackage{amsmath,amssymb,amsfonts}
\usepackage{algorithmic}
\usepackage{multicol}
\usepackage{caption}
\usepackage{graphicx}
\usepackage{textcomp}
\usepackage{xcolor}
\usepackage{csquotes}

\def\BibTeX{{\rm B\kern-.05em{\sc i\kern-.025em b}\kern-.08em
    T\kern-.1667em\lower.7ex\hbox{E}\kern-.125emX}}
\usepackage{hyperref}
\usepackage[space]{grffile}
\usepackage[margin=1.75cm]{geometry}
\usepackage{pdfpages}
\usepackage[capitalize,noabbrev]{cleveref}

\graphicspath{{img/}}

\title{Security and Privacy of Lightning Network Payments with Uncertain Channel Balances}

\begin{document} 


\author[1]{Rene Pickhardt \thanks{rene.m.pickhardt@ntnu.no}}
\author[2]{Sergei Tikhomirov}
\author[2]{Alex Biryukov}
\author[1]{Mariusz Nowostawski}
\affil[1]{Norwegian University of Science and Technology\\
Gj{\o}vik, Norway}

\affil[2]{University of Luxembourg \\
Esch-sur-Alzette, Luxembourg}

\maketitle



\begin{abstract}
The Lightning Network (LN) is a prominent payment channel network aimed at addressing Bitcoin's scalability issues.
Due to the privacy of channel balances, senders cannot reliably choose sufficiently liquid payment paths and resort to a trial-and-error approach, trying multiple paths until one succeeds.
This leaks private information and decreases payment reliability, which harms the user experience.
This work focuses on the reliability and privacy of LN payments.
We create a probabilistic model of the payment process in the LN, accounting for the uncertainty of the channel balances.
This enables us to express payment success probabilities for a given payment amount and a path.
Applying negative Bernoulli trials for single- and multi-part payments allows us to compute the expected number of payment attempts for a given amount, sender, and receiver. 
As a consequence, we analytically derive the optimal number of parts into which one should split a payment to minimize the expected number of attempts.
This methodology allows us to define service level objectives and quantify how much private information leaks to the sender as a side effect of payment attempts.
We propose an optimized path selection algorithm that does not require a protocol upgrade.
Namely, we suggest that nodes prioritize paths that are most likely to succeed while making payment attempts.
A simulation based on the real-world LN topology shows that this method reduces the average number of payment attempts by $20\%$ compared to a baseline algorithm similar to the ones used in practice.
This improvement will increase to $48\%$ if the LN protocol is upgraded to implement the channel rebalancing proposal described in BOLT14.\footnote{\url{https://github.com/lightningnetwork/lightning-rfc/pull/780}.}
\end{abstract}

\section{Introduction}

The demand for public verifiability of Bitcoin, the first decentralized digital currency, limits its throughput to tens of transactions per second.
Second layer, or off-chain protocols are designed to alleviate this issue.
The Lightning Network (LN) is the primary off-chain protocol on top of Bitcoin.
To join the LN, users commit bitcoins on-chain into \textit{payment channels}, which form a network.
LN users can then send bitcoin payments \textit{off-chain}, often utilizing paths of channels (multi-hop payments).
They only have to interact with the Bitcoin protocol again on channel closure.
The LN protocol ensures the atomicity of multi-channel payments and provides an on-chain dispute resolution mechanism.

Information security, defined as a combination of confidentiality, integrity, and availability, is a critical requirement for cryptocurrency-related protocols.
The LN protocol takes deliberate steps to ensure confidentiality (onion routing) and integrity (HMACs).
However, the reliability of LN payments leaves much to be desired.
In the LN context, reliability means that a user with a sufficient balance can quickly send payments with a high probability.

Unfortunately, the design of the LN protocol leads to a lack of payment reliability.
LN payments are usually sent through a path of channels, with the sender specifying the payment path.
The path cannot be changed within one payment attempt, and the order of nodes is fixed due to onion routing (a privacy-preserving measure).
A payment can only be delivered if each channel along the path has sufficient balance to forward it.
However, the balance information of the channel is only known to its owners.
Thus senders often choose paths that are unable to forward the payment and have to make multiple re-tries until a payment finally succeeds.
This makes LN payments unreliable and decreases its speed, especially for larger amounts.
Moreover, trying multiple paths leaks remote channel balances to the sender, eroding LN's privacy, which has also been used in probing attacks~\cite{HerreraJoancomarti2019,Tikhomirov2020}.

Multiple proposals have been discussed to improve reliability in the LN.
One simple but infeasible idea involves sharing channel balances with the network.
This approach is both harmful for privacy and incurs a heavy communication cost, comparable to that of layer-one protocols.\footnote{Communication cost is one of the main bottlenecks to layer-one blockchain scaling -- the exact problem the LN is designed to solve.}
Multi-part payments (MPP) protocols suggest splitting the payment and forward the parts along multiple paths.
MPP has been added to various LN implementations.
Channel rebalancing protocols send payments between different channels controlled by the same user to avoid accumulating balance exclusively on one side of a channel.
Rebalancing needs to be implemented by a significant share of nodes to be effective.
In particular, it is unclear under which conditions splitting payments into parts or rebalancing channels is beneficial.

In this study, we introduce the first probabilistic model of the payment process of the LN that accounts for the uncertainty of channel balances.
We start from modeling a single channel and generalizing to multi-hop and multi-part payments.
The model allows us to express the payment success probability given a path and a payment amount.

We first apply our model to evaluate the reliability of LN payments.
We estimate the expected number of attempts necessary to deliver a payment in single-part and multi-part cases.
Using these expectation values for uniformly distributed channel balances, we create an analytically closed formula that tells us when and into how many parts a single payment should be split.
Similarly, we derive a formula for the expected number of payment attempts $n$ to achieve a given service level objective (i.e.,~deliver a payment with a probability higher than $\sigma$).
In the single-part case, this results in a closed formula $n>\frac{'log(1-\sigma)}{\log(1-s)}$ that depends only on the path success probability $s$ and the service level objective $\sigma$.

Second, we assess one privacy aspect of the LN protocol using our model.
In particular, each payment attempt leaks information to the sender regarding the balances of the channels along the path.
A deliberate technique (probing) to harvest balance information has been described, but we emphasize that a similar information leak also occurs under the normal operation of the LN.
We estimate the information gain (defined as the Kullback Leibler Divergence between the prior and the posterior probability) that a sender achieves depending on payment parameters.
We then discuss the relationship between payment reliability and balance privacy in the LN.\footnote{Calculating information gain requires an assumption about the prior distribution of channel balances. We assume the uniform distribution and will demonstrate later in Chapter~\ref{sec:setup} that this is a reasonable assumption.}

Finally, we apply the model to a more realistic setting.
Assuming a uniform prior distribution of channel balances, we formulate a concrete recommendation for path selection.
In particular, we suggest nodes try paths sorted by their success probability when making payment attempts.
Note that this is a localized measure that any node can adopt, requiring no modifications in the LN protocol.
We develop a simulation framework to evaluate our proposal based on the real-world LN topology, channel capacities, and previously probed channel balances.
We conduct simulated payments of various amounts between $100$~fixed randomly selected sender-receiver pairs.
The results demonstrate that our path selection strategy reduces the expected number of payment attempts by $20\%$.

We envision our model, along with the simulation framework, to be useful for studying the reliability and privacy of the LN under a wide range of assumptions.
In particular, the model may be used for evaluating future protocol modifications.
To exemplify this use case, we study the effect of the channel rebalancing proposal~\cite{BOLT14}.
Our results suggest that rebalancing achieves similar payment reliability as our proposed path selection algorithm even with random path selection.
However, the combination of rebalancing and optimized path selection achieves even better reliability than these measures separately while also significantly reducing the variance of the average number of payment attempts, making the LN more predictable for users.


\section{Background}
\label{sec:background}



\subsection{Channels}

A payment channel allows two parties (Alice and Bob) to re-distribute the initially committed coins.
A channel operates in three stages: opening, operating, and closing.
In the opening stage, the parties establish a \textit{funding transaction} with a 2-of-2 multi-signature output and the first \textit{commitment transaction} representing the initial channel state.\footnote{For each channel state, Alice and Bob have different versions of a commitment transaction with different timelocks imposed on the outputs. This is necessary for the dispute mechanism that disincentivizes malicious closures. These details are irrelevant for our work; therefore, we will assume that each channel state is encoded with one commitment transaction.}
LN channels are single-funded: initially, the full capacity is assigned to the funder.
The value of the 2-of-2 output determines the channel \textit{capacity} and cannot change.
At the operating stage, Alice and Bob update the channel state by co-signing new commitment transactions that are not broadcast until the closing stage.
Each commitment transaction spends the funding transaction's output and creates two or more outputs.
The outputs reflect Alice's and Bob's respective balances and, optionally, in-flight payments.
A channel update may represent a \textit{single-hop} payment from Alice to Bob or be a part of a \textit{multi-hop} payment.
Hashed time-locked contracts (HTLCs) provide atomicity in multi-hop payments.
The LN dispute mechanism ensures that each new commitment transaction invalidates the previous one.
Therefore, exactly one (latest) channel state is valid at any point in time.
In the closing stage, the parties co-sign the final commitment transaction and publish it on-chain.
The LN provides ways to securely close a channel even if one of the parties is offline or malicious, assuming the honest party follows on-chain events and broadcasts dispute transactions if necessary.\footnote{This task may be outsourced to a specialized service -- a watchtower.}
The technical details of these mechanisms are outside of our scope.\footnote{The LN's state invalidation mechanism is economic in nature. A new commitment transaction is signed in a way that enables each party to take all funds from the channel if any of the previous states is published on-chain (i.e.,~a malicious channel closure is attempted).}

\subsection{Payments}

An LN payment is an update of one or multiple channels (single-hop or multi-hop payment, respectively).
To initiate a payment, the receiver generates a random number $r$ and sends its hash $H(r)$ to the sender.
The sender chooses a payment path based on the gossip data and initiates a series of \textit{hash time-locked contracts} (HTLCs) along the path.
An HTLC encodes the following condition within each channel: coins go towards the receiver if it provides a preimage of $H(r)$ before time $t$ or return to the sender otherwise.
Upon receiving an HTLC in the last channel, the ultimate receiver reveals $r$ and triggers balance shifts along the path.
If the receiver does not do so until a specified timeout\footnote{Timeouts differ for different intermediary nodes. These technical details are outside of our scope.}, the payment does not happen.
The same hash value along the path ensures atomicity.

To allow for amounts larger than channel capacity, payments can be split into \textit{parts} (multi-\textit{part} payments, or MPP).
The sender splits the amount $a$ into parts $a_1, \dots, a_k$ and initiates $k$~corresponding payments using the same hash.
The receiver only reveals the preimage upon receiving all parts.\footnote{The receiver can reveal the preimage prematurely, but that would remove the incentive for the sender to send the remaining parts.}
For an MPP to succeed, all parts must succeed.
Different parts of an MPP may be delivered over disjoint paths, partially intersecting paths or even the same path.
MPP has been implemented by major LN implementations: LND~\cite{LND}, c-lightning~\cite{clightning}, and Eclair~\cite{Eclair}.

\subsection{Uncertainty and reliability}

It important to distinguish the channel capacity $c$ and the parties' balances: $b_A$ and $b_B = c - b_A$.
The capacity $c$ is constant throughout the life of the channel.
However, the channel's ability to forward payments depends not only on $c$ but also on $b_A$.
Each channel party cannot send more than it currently has and cannot receive more than its counterparty has.
Channel balances are only known to the channel counterparties.
However, for multi-hop payments, the sender chooses payment paths based only on the publicly available channel graph.
Such paths inevitably include channels with balances unknown to the sender.
This introduces uncertainty: the sender does not know in advance if a path is suitable.\footnote{Of course, the sender only considers channels with capacity $c >= a$ for a payment of amount $a$. This condition is necessary, but not sufficient.}

Balance uncertainty negatively affects payment reliability.
LN payments are made by trial and error: the sender iterates over potentially suitable paths until success.
For multi-part payments, all parts must succeed for the payment to complete.
The sender does not know how many attempts a payment will take and if it will succeed at all.
The number of required attempts in practice translates into waiting time.
Each attempt requires a few seconds, as the LN does not prioritize nodes based on geographical proximity.

\subsection{Privacy and probing}

One may look at the issue of balance uncertainty from the privacy viewpoint.
Outcomes of payment attempts reveal private information about channel balances.
At each failed attempt, the sender receives an error message specifying the failed channel in the path.
This information leak has been exploited to estimate remote channel balances.
In \textit{probing} attacks, a malicious sender issues invalid payments and collects balance information based on error messages~\cite{HerreraJoancomarti2019, Tikhomirov2020, Kappos2020}.
Due to onion routing, such attacks are difficult to curb: an intermediary node only knows the immediately preceding and following nodes in a path.
This makes banning misbehaving nodes challenging, though heuristic-based approaches are possible.

\section{Modeling uncertainty of channel balances}
\label{sec:formalization}

We use probability theory to model the uncertainty of channel balances.
We first consider a single channel.
We then extend our model to payments sent along a path of channels (multi-hop payments).
Finally, we consider multi-part payments (MPP), where the payment amount is split into smaller parts forwarded along different paths.

\subsection{Single-hop payments}
\label{sec:channel}

Let $C$ be a payment channel with capacity $c \in \mathbb{N}$ between nodes $A$ and $B$ who control balances $b_A, b_B \in \mathbb{N}_0$, respectively, such that $b_A + b_B \stackrel{!}{=} c$.
As $c$ is constant, we refer to $b:=b_A$ as the channel balance, since knowing $b_A$ automatically defines $b_B = c - b_A$.
For a given payment channel $C$, we define $\Omega$ to be the discrete probability space of commitment transactions that spend the output of the channel's funding transaction.
Let $X:\Omega\longrightarrow[0,\dots,c]$ be a random variable that maps commitment transactions to balances.\footnote{We do not model pending HTLCs in commitment transactions.}
We choose a probability function $P:2^\Omega \longrightarrow [0,1]$ such that $P(X=b)$ represents the probability that $C$ has the balance $b$.
\footnote{For simplicity, we do not model channel reserves and the inability to create HTLCs for very small (dust) balances.}

A payment from $n_1$ to $n_2$ with the amount $a$ fails if and only if $a > b$.
For a channel $C$ we define the \textbf{channel failure probability} for payment $a$ as:
\begin{equation}
\begin{aligned}
P(X<a) = \sum_{x=0}^{a-1}P(X=x)
\end{aligned}
\end{equation}
Analogously we define the \textbf{channel success probability} as: 
\begin{equation}
\begin{aligned}
P(X \geq a) = 1-P(X<a)
\end{aligned}
\end{equation}



For general probability distributions, $P(X<a+1)\geq P(X<a)$.
Thus, the failure probability is weakly monotonically increasing\footnote{The increase is not necessarily strictly monotonic: depending on the probability function, multiple consequent amounts may have equal failure probabilities.} with the payment amount.


\subsection{Multi-hop payments}
\label{sec:multihop}

An LN payment is usually forwarded through a path consisting of several channels.
Consider payment of amount $a$ forwarded through a path of $l\in\mathbb{N}$ channels with capacities $c_1,\dots,c_l$.
For each channel, we independently select $\Omega_i, P_i$ and $X_i$ as defined previously.
A multi-hop payment succeeds if and only it succeeds at every hop.
Thus the \textbf{path success probability} $s(a)$ for an amount $a$ is the product of channel success probabilities:

\begin{equation}
\begin{aligned}
s(a):=\prod_{i=1}^lP_i(X_i\geq a)
\end{aligned}
\label{eg:path_success}
\end{equation}

The monotonicity of channel success probabilities follows from the monotonicity of the channel success probabilities:
\begin{equation}
\begin{aligned}
a_1 \geq a_2 \Leftrightarrow s(a_1) \leq s(a_2)  
\end{aligned}
\label{eg:path_monotonicity}
\end{equation}

Each factor in the product in Equation~\ref{eg:path_success} is a probability, taking values between $0$ and $1$.
As we only consider channels with balance uncertainty, the factors are strictly smaller than $1$.
Let $p = P_i(X_i\geq a)$ be the largest factor: $p\geq P_j(X_j \geq a)$ for $i\neq j$.
It is trivial that:
\begin{equation}
\begin{aligned}
s(a):=\prod_{i=1}^lP_i(X_i\geq a) \leq p^l
\end{aligned}
\label{eq:path_success_probability}
\end{equation}

Consequently, the \textbf{path failure probability} is expressed as $f(a):=1-s(a)$.
The path success probability decreases exponentially with the number of channels with uncertain balances along the path.
Note that, depending on $P_i$, a longer path may have a higher success probability than a shorter one (for a given amount).

\subsection{Multi-part payments}
\label{sec:multipart}

Multi-part payments (MPP) is a feature of some LN implementations that allows the sender to split a payment of amount $a$ into $k$ parts of amounts $a_1,\dots,a_k$ such that $a=\sum_{i=1}^ka_k$\footnote{It is unclear if the splitting strategies used by LN implementations are optimal with respect to balance uncertainty. In this work, we assume splitting into equal parts. Deriving optimal splitting strategies for MPP remains a question for future research.}.
MPP is motivated by the monotonicity of the path success probability.
For a fixed path, the success probability for each part $s(a_i$) is higher than the success probability $s(a)$ for the full amount.
In reality, partial payments travel along $k$ different - but not necessarily disjoint - paths, resulting in $k$ different success probabilities $s_1(a_i),\dots,s_k(a_k)$.
The success probability of a multi-part payment is computed as $\prod_{i=1}^ks_i(a_i)$.
In general, we can not say if it is higher, lower, or equal to the success probability $s(a)$ for the full amount in a single-part payment.

In summary, within the model, the usefulness of multi-part payments depends on three aspects:
\begin{enumerate}
\item the full amount $a$;
\item the way how $a$ is split into parts $a_1,\dots,a_k$;
\item path success probabilities $s_i(a_i),\dots,s_k(a_k)$ for the paths that the partial payments take.
\end{enumerate}

\section{Applications of the model}
\label{sec:application}

In the previous section, we have introduced a model of channel balance uncertainty for multi-hop and multi-part payments.
Our model has shown that payments succeed more often with smaller amounts and shorter paths.
We now use the model to explore three specific questions of interest: estimating the expected number of payment attempts, establishing service level objectives, and estimating privacy leaks.

\subsection{Expectation number of payment attempts}

According to the LN protocol, sending a payment may involve multiple failed attempts until a payment is successfully delivered.
We can express the expected number of payment attempts using our model.
The probability $p$ for a payment of amount $a$ to only be successful at the $n$-th attempt after $n-1$ unsuccessful attempts is:

\begin{equation}
\begin{aligned}
p(a,n)=s_n(a)\prod_{i=1}^{n-1}f_i(a)
\end{aligned}
\label{eq:attempts}
\end{equation}

Assuming a fixed path success probability $s = s_1\dots,s_n$ for all attempts, and that attempts are independent, we can use Bernoulli trials:
\begin{equation}
\begin{aligned}
B(s;n,k) = {n \choose k}s^k(1-s)^{n-k}
\end{aligned}
\label{eq:bernoulli}
\end{equation}

In the single-part case ($k=1$), just one attempt must be successful after $n-1$ failed attempts.
Thus $p(a,n)=s(a)\cdot B(s;n-1,0)$, which is exactly Equation~\ref{eq:attempts}.
Similarly, in the multi-part case, $k$~parts must succeed at the $n$-th attempt after having $k-1$ successes after the preceding $n-1$ attempts, resulting in $s(a_k)\cdot B(s;n-1,k-1)$.
If $n$ is the random variable representing the number of attempts, we can compute the expectation value $E[n]$.

According to the theory of negative Bernoulli trials~\cite{Papoulis1991}, for an arbitrary value of $k$, the expectation value is:

\begin{equation}
\begin{aligned}
E[n]=\frac{k}{s}
\end{aligned}
\label{eq:expectation_value}
\end{equation}

For $k=1$, the formula simplifies to: $E[n]=\frac{1}{s}$.

Recalling the monotonicity of the path success probability, we see that smaller amounts lead to higher values of $s(a)$ and thus to a lower expected number of attempts.
However, we stress that Equation~\ref{eq:expectation_value} does not imply that splitting a payment always decreases the expected number of attempts.
Indeed, for a fixed path success probability $s$, we have $\frac{k_2}{s} > \frac{k_1}{s}$ if $k_2>k_1$.
However, $s$ increases for larger values of $k$.
Therefore, we cannot claim that multi-part payments yield a lower expected number of attempts in the general case.
We need to consider channel success probabilities to find out under which conditions multi-part payments are beneficial.

\begin{figure}[h!]
  \includegraphics[width=0.5\textwidth]{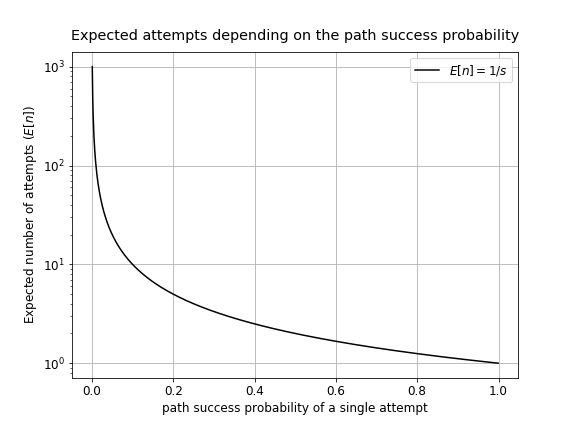}
\captionof{figure}{Expected number of payment attempts for a single-part payment depending on the success probability of a single attempt.}
  \label{fig:expectationValue}
\end{figure}

Figure \ref{fig:expectationValue} shows the expected number of attempts for $k=1$ depending on the success probability $s$ of a single payment attempt.
We observe that lower values of $s$ increase the expected number of attempts.
For instance, decreasing $s$ from $50\%$ to $10\%$ increases the expected number of attempts from $2$ to $10$.

Again, we emphasize that this result depends on the assumption that $s$ is fixed for a given amount for all paths from the sender to the receiver.
This assumption is likely unrealistic.
However, the result may still help nodes select paths.
Assuming there is a way to estimate the path success probability before sending the payment, the sender could make statements like:
\begin{displayquote}
If I only choose paths with a success probability of more than $25\%$, the average number of required attempts is expected to be lower than four.
\end{displayquote}
Of course, the sender could choose different values for the path success probability or for its target expected number of attempts.

Equation~\ref{eq:expectation_value} confirms that high path success probabilities (coming from small amounts and short paths) yield a low number of necessary payment attempts.

\subsection{Establishing service level objectives}

Expectation values do not convey information about their underlying probability distribution.
In particular, the expected number $n$ of attempts does not imply a high likelihood of completing a payment after $n$ attempts or fewer.

Thus we now pose the question: how many attempts are necessary to guarantee a payment success probability $\sigma$?
We call $\sigma$ our service level objective.
We start again from Formula~\ref{eq:bernoulli} and study the case of a single-part payment.
In this case, at least one of the $n$ attempts must be successful to complete the payment.
Thus we search for $n$ such that:
\begin{equation}
  \sum_{k=1}^nB(s;n,k) > \sigma
\end{equation}
Since $\sum_{k=0}^nB(s;n,k)\stackrel{!}{=}1$ for all $n$, we can simplify this inequality to $1-B(s;n,0) > \sigma$.
The probability of having at least $1$~successful attempt after $n$ attempts is the same as $1$ minus the probability of having exactly $0$ successful attempts.
As $B(s;n,0) = (1-s)^n$, the inequality simplifies to:
\begin{equation}
\begin{aligned}
1-(1-s)^n & > \sigma \\
  \Leftrightarrow 1-\sigma & > (1-s)^n \\
  \Leftrightarrow \log(1-\sigma) & > n\log(1-s) \\
  \Leftrightarrow \frac{\log(1-\sigma)}{\log(1-s)} & < n
\end{aligned}
\label{eq:SLO}
\end{equation}

\begin{figure}[h!]
\includegraphics[width=0.5\textwidth]{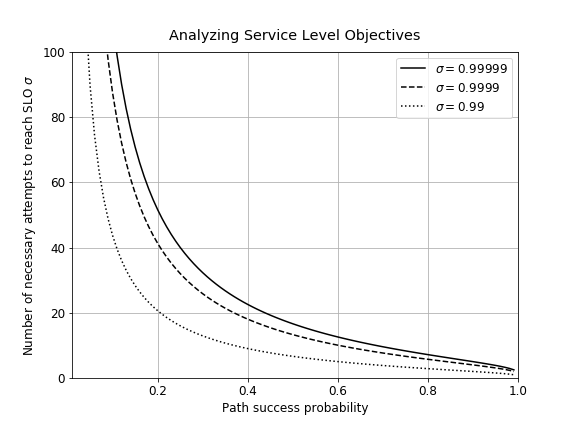}
\captionof{figure}{The number of attempts to achieve a service level objective depending on the success probability of a single attempt.}
\label{fig:SLA}
\end{figure}

Thus the number of required attempts to achieve a given service level objective is reversely proportional in $\log(1-s)$.
We see clearly that higher service level objectives require more payment attempts for a fixed success rate of a single attempt (Figure~\ref{fig:SLA}).

\begin{figure}[h!]
\includegraphics[width=0.5\textwidth]{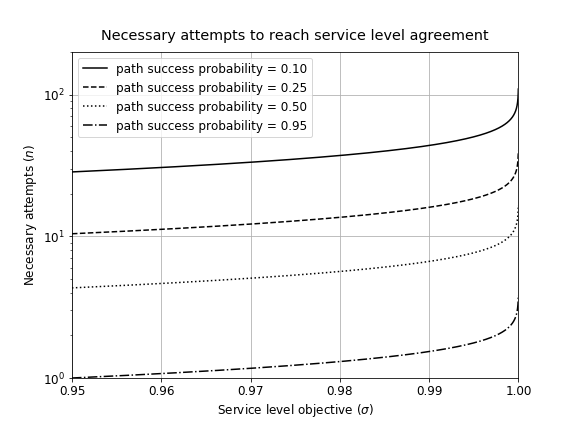}
\captionof{figure}{The number of attempts necessary to meet a service level objective for various path success probabilities.}
\label{fig:SLA_sigma}
\end{figure}

Consider this function with the service level objective $\sigma$ as the depending variable in the range of high service level objectives (Figure~\ref{fig:SLA_sigma}).
We observe that increasing the path success probability decreases the number of attempts to achieve even higher service level objectives.
For instance, increasing the path success probability from $10\%$ to $50\%$ decreases the number of attempts to reach $\sigma=0.99$ from $40$ to $7$.

Due to the fact that the logarithm declines quickly when getting close to 0, we see that one should focus on increasing path and channel success probabilities.

We can also use Bernoulli trials to make statements about the number of attempts for multi-part payments.
Assuming $s_{mpp} = s_i(a_i) \forall i$, we can express the probability for at least $k$ partial payments to succeed as:

\begin{equation}
1 - \sum_{i=0}^kB(s_{mpp};n,i) > \sigma
\end{equation}

We are unaware of an analytic solution for $n$ in the general case of arbitrary values of $k$.
However, choosing a prior distribution for $s$ allows us to solve it numerically, as we will do in Section \ref{sec:example}.

\subsection{Privacy and information gain}
As channel balances are unknown, the fact that the sender can observe events like the success or failure of a payment attempt leaks information.
This observation has been used in the past to probe the network for channel balances. 
We can use our framework to quantify the information gain that the sender yields from successful or failed payment attempts.

Modeling the balance of a payment channel with capacity $c$, we can measure the uncertainty of the balance by computing the entropy~\cite{shannon1948mathematical} $H$ using a probability function $P$ as:
\begin{equation}
  H(P) = - \sum_{x=0}^c P(X = x)\cdot\log(P(X = x))
\end{equation}

For a payment amount $a$, we can define the event $A:=X<a = \{c \in \Omega | X(c) < a\}$.
In the case of event $A$, the payment attempt fails.
Therefore, if we tried to send the amount $a$ and have failed, we can update the channel balance probability by switching to the conditional probability.

\begin{equation}
  P(X = b|A)=\frac{P(X^{-1}(b) \cap A)}{P(A)}
\end{equation}
If $b\notin A$, the intersection is the empty set, and the probability $P(X=b|A)=0$ as it should be. 
Thus the conditional probability can be expressed as:
\begin{equation}
P(X=b|A) =
\begin{cases}
  \frac{P(X=b)}{P(A)} & \text{for } b < a \\
  0 & \text{else}
\end{cases}
\end{equation}

This updated probability can be used to compute the information gain as the Kullback Leibler divergence between the prior probability $P(X)$ and the posteriori probability $P(X|A)$:
\begin{equation}
\delta = \sum_{x=0}^cP(X=x|A)\cdot\log\left(\frac{P(X=x|A)}{P(X=x)}\right)
\label{eq:information_gain}
\end{equation}

Similarly, we can study the successful case by calculating the conditional probability on the event $X \geq a$.
We can also compute the information gain for a multi-hop payment by adding up the information gains of all channels in the path.

\section{Theoretic results with uniform priors}
\label{sec:example}
The computation of the expectation values, service level objectives, and information gains depends on selecting a probability distribution for channel balances.
We will now apply our model, assuming a uniform distribution for the channel balance probability.\footnote{In Chapter\ref{sec:setup}, we will show that uniform distribution is a reasonable choice to describe the Lightning Network. Moreover, uniform distribution allows us to simplify the equations, which is the main reason why we use it to exemplify the application of the model.}
For a channel of capacity $c$, we can select a uniform prior distribution $P$ and random variable $X$ such that $P(X=b) = \frac{1}{c+1}$ is constant for any balance value $b$.
The channel failure probability $P(X < a)$ thus equals $\sum_{x=0}^{a-1}P(X=x) = \sum_{x=0}^{a-1}\frac{1}{c+1} = \frac{a}{c+1}$.
The failure probability is proportional to the payment amount and indeed increasing with larger amounts.
Similarly, the channel success probability $P(X\geq a)=\frac{c+1-a}{c+1}$ is increasing with smaller amounts.
Note that the success probability also increases with larger channel capacities -- a statement we cannot make for an arbitrary probability distribution.

For multi-hop payments, we can compute the success probability for amount $a$ as the product of all channel success probabilities: $s(a) = \prod_{i=1}^lP_i(X_i>a)$.
For simplicity, in the rest of this section, we assume that all channels are of equal capacity.
We may interpret the results as a worst-case scenario if we choose $c = \min\{c_1,\dots,c_l\}$.
In that case, $P(X>a) = \frac{c+1 - a}{c+1}$ for every channel, resulting in $s(a)=\left(\frac{c+1-a}{c+1}\right)^l$.

\begin{figure}
\includegraphics[width=0.5\textwidth]{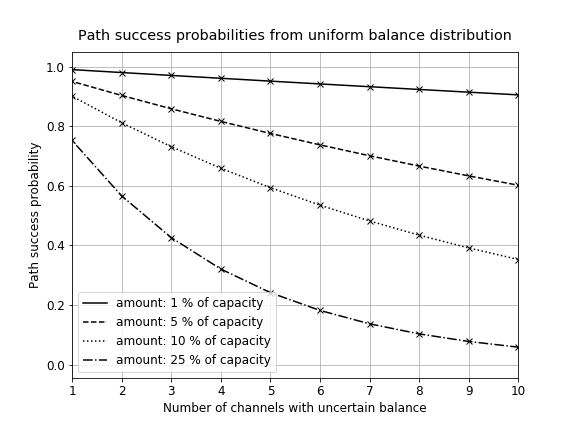}
\captionof{figure}{Success probability depending on path length, for selected payment amounts (constant-capacity channels, uniform balances).}
\label{fig:uniform_path_success_probability}
\end{figure}

For each of the selected values of $a$, the path success probability declines exponentially with the path length (Figure~\ref{fig:uniform_path_success_probability}).
(Recall that we define path length as the number of channels with uncertain balances in the path, i.e.,~we do not count the first channel.)

We also observe that success probability is higher for small amounts.
For example, sending $10\%$ of the (constant) channel capacity along a path with $4$~unknown channels has just $65.9\%$ chance to succeed.

\begin{figure}
\includegraphics[width=0.5\textwidth]{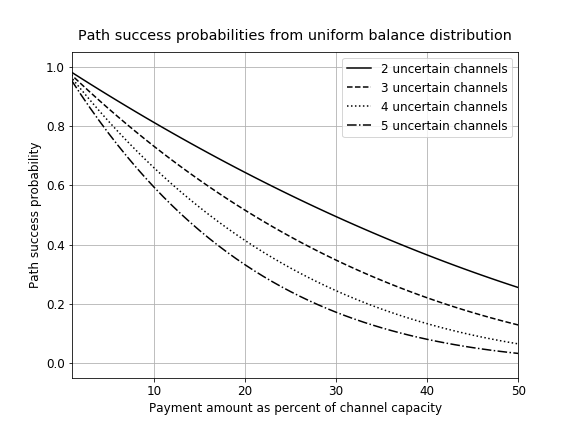}
\captionof{figure}{Success probability depending on payment amount, for selected path lengths (constant-capacity channels, uniform balances).}
\label{fig:uniform_path_success_probability_2}
\end{figure}

Flipping the variables, we observe that payment success probabilities decline exponentially with the growth of the payment amount (Figure~\ref{fig:uniform_path_success_probability_2}).

\begin{figure}
\includegraphics[width=0.5\textwidth]{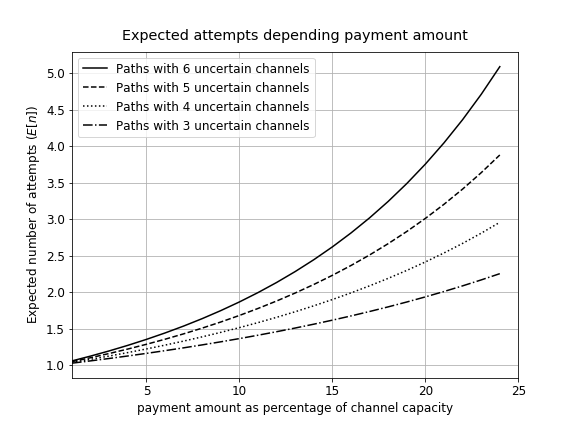}
\captionof{figure}{Expected number of payment attempts depending on payment amount, for various path lengths (constant-capacity channels, uniform balances).}
\label{fig:expectation_uniform}
\end{figure}

Depending on the payment success probability, we can compute the expected number of attempts necessary for a successful payment.
Assuming constant channel capacities and uniformly distributed balances, the success rate depends only on the payment amount and the number of uncertain channels.
We plot the expectation values for various path lengths (Figure~\ref{fig:expectation_uniform}).
Note the difference compared to calculations in Section~\ref{sec:application} (Figure~\ref{fig:expectationValue}) that only showed how the expected number of attempts depends on the success probability of a single attempt and not on a payment amount.
As shorter paths and smaller amounts yield higher payment success probabilities, we observe a lower expected number of payment attempts in such cases.

In particular, all values on the horizontal lines in Figure~\ref{fig:expectation_uniform} result in the same success probability and thus the same expected number of attempts.

Consider a sender who wants their payment of amount $a$ to succeed after two attempts on paths with up to $3$~uncertain channels.
These requirements are satisfied if $a$ does not exceed $20\%$~of the channel capacity $c$.
Therefore, the sender may reason as follows:
\begin{displayquote}
  To send the amount $a$ using $2$ attempts, I should only consider paths with $3$~routing nodes or fewer, where every channel has a capacity of at least $5 \cdot a$ (the payment amount should not exceed $0.2\cdot c$, thus $0.2\cdot c = a \Leftrightarrow c = 5\cdot a$).
\end{displayquote}

Recall that the expected number of attempts for a multi-part payment of $k$ parts is $E[n]=k/s$.
Larger values of $k$ require more attempts for a fixed value of $s$.
However, this does not imply that splitting a payment is never beneficial, as $s$ depends on the amounts.
Assume a split of the amount $a$ into $k$ equal parts of $\frac{a}{k}$.
We can derive the optimal number of parts to achieve the lowest expected number of attempts. 

\begin{figure}
\includegraphics[width=0.5\textwidth]{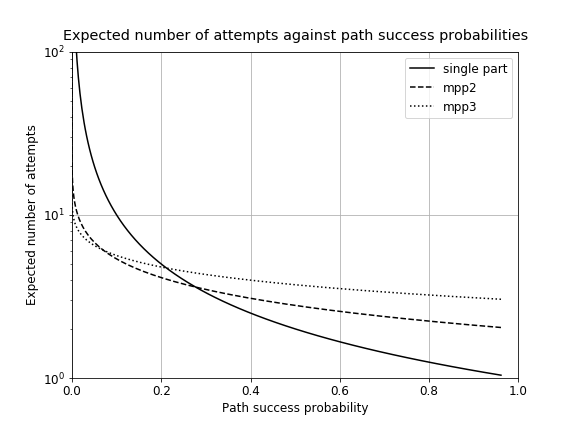}
\captionof{figure}{Expected number of attempts depending on the success probability for the full amount, for selected numbers of payment parts.}
\label{fig:expectation_mpp}
\end{figure}

For a given $a$, we compute $s(a)$ as the success probability of that amount.
We then compute the expected number of attempts as $E[n] = \frac{k}{s\left(\frac{a}{k}\right)}$.
Considering $E[n]$ for $k=1, 2, 3$, we observe the curves of similar shapes (Figure~\ref{fig:expectation_mpp}).
The curves for higher values of $k$ are shifted to the left.
For each path success probability $s(a)$, the sender should split the payment according to the lowest curve, minimizing the expected number of attempts.
Note that those curves are the theoretic lower bound of what is possible on average.
As we move from higher to lower values of $s(a)$, we encounter break-even points (curve intersections), where it becomes beneficial to split the payment into more parts.

Higher amounts yield lower path success probabilities $s(a)$.
With $s(a) < 0.3$, the MPP with two parts starts to show a lower number of expected payment attempts.
This means that, at a certain payment amount, it is reasonable to split it into $2$~parts.
If the amount rises and the path success probability drops, the next break-even point is at $s(a) < 0.1$: for lower path success probabilities, the payment should be split into $3$~parts.

The intersection points can be computed analytically by solving the following equation for $a$:

\begin{equation}
\frac{k_1}{s\left(\frac{a}{k_1}\right)} = \frac{k_2}{s\left(\frac{a}{k_2}\right)}
\end{equation}

In the case of a uniform \textit{channel} success probability, this results in:

\begin{equation}
  a = c\cdot\frac{\left(1-\sqrt[^l]{\frac{k_2}{k_1}}\right)}{\left(\frac{1}{k_2}-\frac{\sqrt[^l]{\frac{k_2}{k_1}}}{k_1}\right)}
  \label{eq:mpp_split_points}
\end{equation}

Considering paths with $l=2$ routing hops, $k_2=4$, and $k_1=1$, we get $a=c\cdot\frac{4}{7}$.
This means that for a payment amount higher than $\frac{4}{7}$ of the channel capacity, a multi-part payment with $4$ equal-sized parts requires fewer attempts than a single-part payment.
In general, the sender can derive the optimal number of parts by solving this equation for $k_2=k_1+1$ for all values of $k_1$ until the desired payment amount is reached.
Recall that to deliver a multi-part payment with $k$ parts with a probability of at least $\sigma$, we need to preform $n$~attempts, where $n$ is such that $1 - \sum_{i=0}^{k}B(s,n,i) > \sigma$.
To the best of our knowledge, no analytic formula exists to solve this equation in the general case.\footnote{\url{https://math.stackexchange.com/questions/232464/how-many-tries-to-get-at-least-k-successes}} 
We solve this equation numerically for $n$.

\begin{figure}
\includegraphics[width=0.5\textwidth]{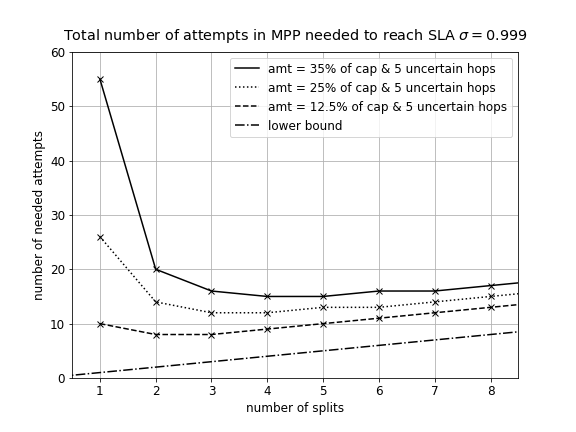}
\captionof{figure}{The required number of attempts to reach a service level objective of $0.999$ depending on the number of payment parts, for various amounts and path lengths (constant-capacity channels, uniform balances).}
\label{fig:mpp_attempts}
\end{figure}

The numerical solution can be found efficiently with brute force after making the following observation.
The number of necessary attempts for a single-part payment is known and is computed as $n$.
In that case, we must not split the payment into more than $n$ parts, as we would need more than $n$ attempts to deliver all of them.\footnote{This assumption can actually be discussed and depends on our goals. If the sender wants to optimize for the time required for the payment completion, multi-part payments may be beneficial outside of these bounds as they can be executed concurrently.}
For each $k=1,\dots,n-1$, we compute the number of necessary attempts to reach the target success probability $\sigma$.
If for the current value of $k$ the number of attempts exceeds $n$, we abort the computation.
Using this method, we derive Figure~\ref{fig:mpp_attempts}.

The dashed line $y=k$ shows the absolute lower bound for the number of attempts (we need at least as many attempts as parts).
We observe that splitting the payment into $2$ or $3$ parts seems almost always preferable.
However, while the success probability of even smaller amounts increases, the exponential dependency of the number of hops starts to dominate.
We conclude that multi-part payments improve payment reliability but only up to a certain extent.
There is a lower bound of what can be achieved with multi-part payments.

Assuming a uniform prior balance distribution $P$, we can compute the entropy: H(P)=$\log(c+1)$. \footnote{The proof of this can be found in Appendix~\ref{appendix:proofs}.}
As before, the event $X < a$ denotes a failed payment attempt for amount $a$.
The event $X<a$ means that the channel balance is lower than $a$; therefore, the probability mass $P(X\geq a)$ is redistributed.
A uniform distribution implies that the channel balance is equally likely to take any value $0,\dots,a-1$ with probability $\frac{1}{a}$.
The entropy of the conditional probability is $\log(a)$.
The information gain can thus be computed as $\delta = \log(c+1)-\log(a)$.
Analogously, in case when $a$ is successfully forwarded, the entropy is $\log(c-a+1)$, and the information gain is $\delta = \log(c+1) - \log(c-a+1)$.
In this way, we can measure how much the sender learns from payment attempts for both single- and multi-part payments.
Note that in a practical setting, one needs to account for the fact that paths may share channels. Information gain for repeatedly used channels must only be counted once.

\section{Experimental setup}
\label{sec:setup}

Our next goal is to confirm the applicability of the model.
Towards this end, we develop a simulation framework.
The framework takes as input a channel graph populated with channel capacities and, optionally, balances.
The simulator supports two modes of operation regarding balance generation: static and dynamic.
In the static mode, balances are fixed in the channel graph (they may be pre-generated according to a given distribution or obtained from the real network).
While delivering a payment, the simulator keeps track of failed channels and excludes them during the retries.
In the dynamic mode, balances are generated at runtime for each payment attempt according to a given distribution.
Therefore, re-sending a payment through a failed channel may still succeed.
The static mode closely reflects the real LN, while the dynamic mode matches our theoretical model.

Given a channel graph, a mode of operation (static or dynamic), and a prior distribution (if balances are unknown), the simulator generates random sender-receiver pairs and conducts payment attempts until success.
The simulator supports MPP for equally-sized parts: for a $k$-part payment to succeed, $k$~partial attempts must succeed.
For each payment attempt, the simulator stores whether it is successful, the theoretical path success probabilities, and the information gain.

Balance generation requires making an assumption on the prior distribution of channel balances.
We conduct a measurement of the real LN using a probing tool to estimate the real balances of a small subset of channels.
We probe $500$~randomly selected channels, or $4.3\%$~of live and active channels as of October~2020.
We observe a mixed distribution of the channel balances: channels may be divided into two groups with bimodal and uniform balance distribution.
In the bimodal group, which consisted of roughly $30\%$ of the channels, most of the capacity is on one side of the channel.
Other channels exhibit a distribution close to uniform, i.e.,~balance values between $0$ and channel capacity seem equally likely.
A repeated probing of $500$ channels with another channel selection method (weighted by centrality instead of uniformly at random) yields similar results.
We conclude that the observed balance distribution is characteristic of the LN and not explained by our channel sample.

Next, we can make some assumptions regarding the level of activity of channels.
LN channels are single-funded: initially, the full capacity is on one side.
The channels from the bimodal group are therefore likely to have rarely or never been used. 
We conclude that most payments flow through the uniform group of channels.
We used our probed samples to predict $137$~nodes, which would mainly have active channels.
Indeed $882$~channels across those nodes exist - $101$ of which we had already probed.
In a final crawl, we probed the remaining $781$~channels and observed that in this subnetwork, the balance distribution was close to uniform as can be seen in Figure~\ref{fig:rebalanced} later in Section~\ref{sec:results}. 
In what follows, we use a sub-graph of the LN consisting of those $137$~nodes and $882$~channels.
We refer to this sub-graph as the \textit{active kernel}.
For experiments that do not involve real balance data, we generate balance values following a uniform distribution.
This means that for a channel with capacity $c$ every balance value $b$ between $0$ and $c$ has the same probability $\frac{1}{c+1}$.

%

\section{Results}
\label{sec:results}

We now aim to answer three questions:
\begin{enumerate}
\item Does the simulation framework under specific conditions reproduce the model?
\item What practical recommendations can we give based on the model, and how useful are they?
\item What is the impact of future protocol changes?
\end{enumerate}

\subsection{Testing the uniform model}

\begin{figure}
\includegraphics[width=0.5\textwidth]{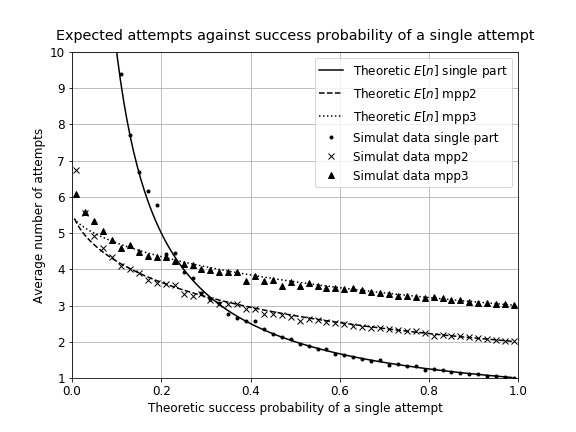}
\captionof{figure}{Expected number of attempts from the model and the simulation for payments of $1$, $2$, and $3$ parts. The expected number of attempts measured in the simulation closely matches the theory.}
\label{fig:expectation_simulation}
\end{figure}

First, we check that our simulator implements the model correctly.
To that end, we consider the topology of the active kernel of the LN and paths of lengths from $1$ hop to $5$ hops.
We set the capacities of all channels to constant values and select payments amounts of $1\%,2\%,\dots,99\%$ of the channel capacity.
This setup closely resembles the uniform model as described in Section~\ref{sec:example}.
As the model assumes independent trials, we generate balances dynamically: at each payment attempt and each channel in the path, we choose balances of channels along the path uniformly at random between $0$ and the channel capacity.
For each path, we compute the theoretic path success probability (Equation~\ref{eq:path_success_probability}).
The expected number of payment attempts given a path success probability in the simulation experiment match closely the predicted value from the model (Figure~\ref{fig:expectation_simulation}).

Thus the theoretic model is confirmed by our simulation framework.
We recall that according to Formula~\ref{eq:mpp_split_points}, the break-even amounts for optimal splits into multiple parts depend on the number $l$ of uncertain channels that payments take (Figure~\ref{fig:expectation_mpp}).
In the simulation, the used paths contained $1.5$~channels with unknown balances on average.
We thus used the same value for $l$ to compute the theoretic expectation values which we used to compare the simulated data against.

\begin{figure}
\includegraphics[width=0.5\textwidth]{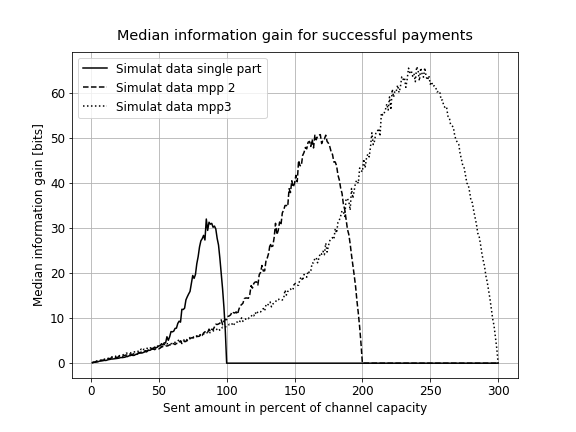}
\captionof{figure}{Information gain for payments split into $1$, $2$, and $3$ parts. Most information is leaked during payments of around $80\%$~of channel capacity.
}
\label{fig:entropy_mpp}
\end{figure}

We then measure the information gain that the sender yields while completing a payment.
We consider amounts from $1\%$ to $300\%$ of the channel capacities in $1\%$ steps, to also test the MPP case.
We randomly select $100$~payment pairs and make payment attempts until success.
For each attempt, we calculate the information gain as per Equation~\ref{eq:information_gain}.
Figure~\ref{fig:entropy_mpp} depicts median values across the payment pairs for each amount.

We see that larger amounts generally lead to the sender obtaining more information before payment succeeds.
Indeed, larger amounts need more attempts, and each attempt leaks information.
For small amounts, information gain shows slow linear growth at low amounts.
At around half of channel capacity, it starts growing much faster.
We explain it as follows: higher amounts yield higher failure rates and thus gain more information (the experiment is conducted until success).
However, the information gain quickly drops to zero when amounts approach the channel capacity: failures in such a setting is "expected" and only yields a small amount of information.
Multi-part payments exhibit a similar picture scaled accordingly.

\begin{figure}
\includegraphics[width=0.5\textwidth]{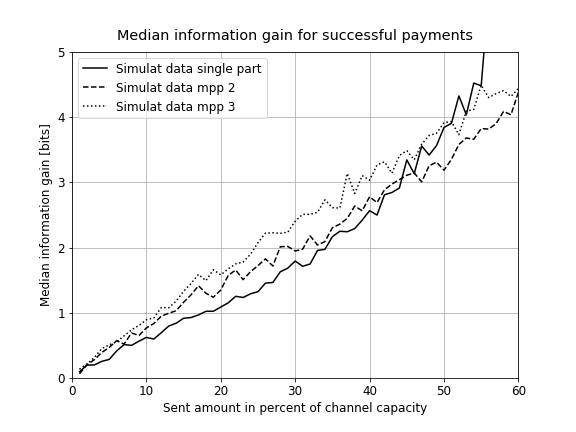}
\captionof{figure}{Closer look at the information gain for small payment amounts}
\label{fig:entropy_mpp_zoomed}
\end{figure}

Figure~\ref{fig:entropy_mpp_zoomed} is an enlarged part of Figure~\ref{fig:entropy_mpp} focusing on small amounts.

The information gain increases with the number of payment parts but is roughly the same for a different number of parts.
If the sender wants to complete the payment using the least amount of collected balance information, its strategy is similar to what we described earlier with respect to the expected number of payment attempts.
For small amounts, one should not split the payment.
Then, for amounts above a certain threshold (around $45\%$ in Figure~\ref{fig:entropy_mpp_zoomed}, one should split a payment into $2$ parts, and after another threshold (around $74\%$) it is best to split it into $3$ parts.

\subsection{Testing an improved path selection algorithm}

Based on our theoretical findings, we now propose an optimized path selection algorithm for better payment efficiency.
We suggest iterating through paths sorted by pre-computed success probability.
We test this proposal using the model in a more realistic setting.
We use the channel capacities from gossip data and balances of the active kernel of the LN obtained using a probing tool.

Recall that a sender in the LN needs to deliver a given amount and iterates through a list of potentially suitable paths until one succeeds.
Note that the paths may have different success probabilities, according to our model.
We estimate the average number of attempts for a successful payment along a path of a given success probability.

In most cases, the data is too sparse to get meaningful results.
The situation becomes worse in the multi-part case where payment parts go through paths with different success probabilities, which prevents us from making a meaningful comparison of expectation values.

We thus adopt the simulator to test the outcomes of the model via the following experiment.
We select $100$~random payment pairs and try to send amounts from $1$ to $20$ mBTC. 
We try to deliver the amount with two different path selection strategies:

\begin{enumerate}
\item \textbf{Baseline}: We randomly select paths between the sender and the receiver and try one after another until we succeed. The paths are ordered by length, which makes this strategy similar to the ones implemented in LN clients.
\item \textbf{Maximum Likelihood}: We pre-compute $1000$~shortest paths iterate through them from the highest path success probability to the lowest.
\end{enumerate}

Mimicking the behavior of real LN nodes, we remember failed channels and do not re-try paths that contain them (for both strategies).
Similarly, we only try paths for which the first channel has sufficient balance: in the real LN, this information is known to senders, and they would not try to make a payment they know will fail.\footnote{We ignore the fact that on the real LN, the receiving nodes can also indicate in the invoice message on which channel they can receive the requested amount.}
Both these optimizations can be expressed within our model as follows.
Excluding failed channels corresponds to setting the channel probability to $0$.
Similarly, first hops receive a success probability of $0$ if the balance is insufficient or $1$ otherwise.
For the next attempts, paths that include hops with success probability $0$ are excluded.\footnote{An even better strategy would be to also re-calculate success probabilities accounting for successful hops that now have probability $1$. We leave the implementation and evaluation of this method for future work.}

\begin{figure}
\includegraphics[width=0.5\textwidth]{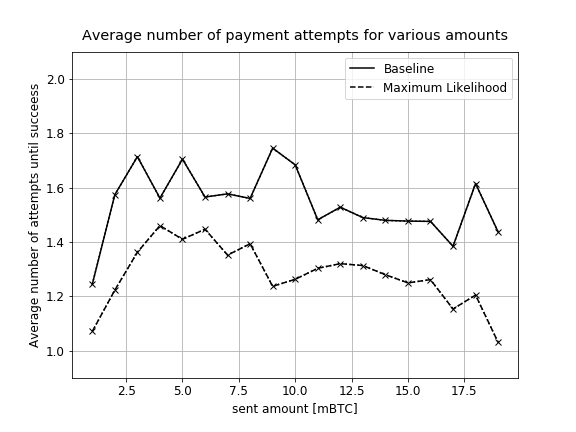}
\captionof{figure}{Average attempts for the baseline and optimized path selection strategies (simulated based on real-world LN data).}
\label{fig:average_attempts_real_ln}
\end{figure}

Figure~\ref{fig:average_attempts_real_ln} shows the distribution of the number of necessary attempts for the two strategies.

The optimized strategy yields fewer attempts on average than the baseline strategy for all considered amounts.
Across all amounts, the number of attempts drops by $20\%$: from $1.54$ to $1.28$.
This result generalizes to multi-part payments (not depicted) with the caveat that a $k$-part payment requires at least $k$~attempts to succeed.
The optimized path selection strategy is beneficial for amounts where MPP should be applied as well.
We emphasize that the optimized path selection strategy does not require protocol modifications and can be straightforwardly added to existing LN implementations.

We note that in the theoretical model, we control the expectation value for a fixed path success probability.
Because of the various channel capacities and the topology of the real Lightning Network, there are hardly any paths with the same success probability for a given amount and payment pair.
Thus it becomes infeasible to compute the average number of attempts for a fixed success probability and compare them to the theoretic expectation value.
Even though we control for the amount, the measured average number of attempts in both the baseline case and the maximum likelihood case is computed from paths with different success probabilities.

\subsection{Testing potential protocol changes}
\label{sec:testing-rebalancing}

\begin{figure}
\includegraphics[width=0.5\textwidth]{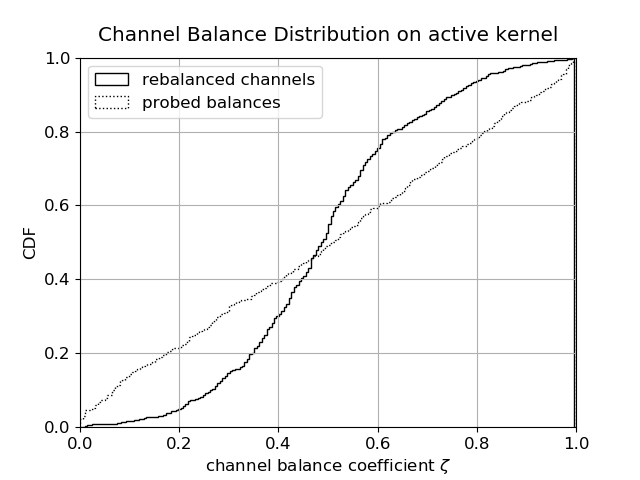}
\captionof{figure}{Rebalancing payment channels according to \cite{} on a real LN Snapshot with real capacities and balances shift the shape of the balance distribution from uniform to normal.}
\label{fig:rebalanced}
\end{figure}

The proposed framework can also be used to test the feasibility of potential protocol changes.
For example, BOLT14~\cite{BOLT14} proposes a channel rebalancing protocol for the LN.
This proposal is motivated by \cite{Pickhardt2019a} that predicts rebalancing to increase the success rates of single-satoshi payments and median possible payment amounts.
We apply the proposed rebalancing algorithm on our LN snapshot.
As a result, the channel balance distribution shifts from uniform-shaped to normally-shaped (Figure~\ref{fig:rebalanced}).

\begin{figure}
\includegraphics[width=0.5\textwidth]{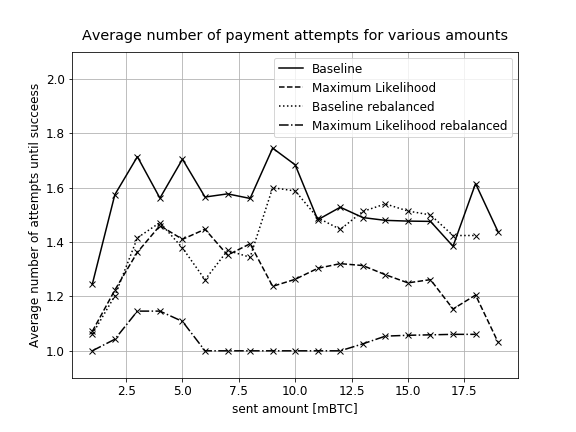}
\captionof{figure}{Comparing the average number of attempts for the probed and the rebalanced network with both path selection strategies.}
\label{fig:average_attempts_rebalanced}
\end{figure}

According to our model, we expect lower average payment attempts for lower amounts as the channel success probabilities increase in the case of normal distributions.
However, this comes at the cost of even higher failure rates for larger payments.
We study the effects of rebalanced channels in Figure~\ref{fig:average_attempts_rebalanced}.

For the baseline strategy, the rebalanced network outperforms the non-rebalanced one for smaller amounts but performs equally poorly for larger amounts. \footnote{At even larger amounts, we would see that rebalancing performs worse in the baseline case.}
However, when also applying the optimized path selection strategy, the rebalanced network yields a consistently lower average number of attempts than the non-rebalanced one.
We conclude that the best-known method to achieve high payment reliability in the LN is the combination of MPP with an optimally selected number of parts, optimized path selection, and channel rebalancing.

\section{Discussion, limitations, and future work}
\label{sec:discussion}

Uncertain channel balances are not the only reason why LN payments are unreliable.
Payments may also fail due to attacks where a malicious routing node delays forwarding~\cite{cryptoeprint:2019:1149} or floods channels with small payments~\cite{harris2020flood}.
In the case of a protocol breach, channels might have to be resolved on-chain, imposing long timelocks until in-flight payments resolve.
Modeling these sources of unreliability remains a topic for future research.

The main limitation of our model is that its application in a realistic setting requires the assumptions of a prior balance probability distribution.
As we have observed in Section~\ref{sec:testing-rebalancing}, switching from a uniform to a normal distribution heavily changes the outcomes.
Thus users of the proposed optimized path selection strategy should use a realistic prior distribution to compute theoretic success probabilities.
To keep an up-to-date estimation of the balance distribution, a node can either regularly probe a few channels or rely on a trusted party that shares its probing results.

A related limitation is that our current model and evaluation only partially capture the temporal dynamics of the LN.
All our test data points have been collected by assuming a single payment on a static LN snapshot.
We did not simulate the effectiveness of our optimized path selection algorithm by conducting a series of payments across different payment pairs, which, if delivered successfully, change balance values.
Such a simulation is challenging because of the lack of available transaction data for the LN, and applying data sets from other payment networks to the LN context is methodologically difficult and remains a question for future work.
Application of our model in the dynamic setting requires an assumption that the balance distribution does not change significantly, which may be tested by periodically probing balances.

Another limitation is that for practical and ethical reasons, we have conducted our experiments on a rather small subgraph of the LN.
Thus the number of expected average numbers of attempts in our simulation tends to be small.
However, we believe that the model can be applied to improve payment reliability for the full LN snapshot.

Our closed formula and experiments regarding the optimal MPP splits only consider splits into equal-sized parts.
Practically speaking, MPP splitting should optimize for the overall success probability.
A completely different type of splitting strategies would split payments in a way that rebalances the sender's own channels.
This would trade some reliability to help the overall network to approach normal balance distribution.
Deriving optimal splitting strategies for various goals is an interesting line for future research.

\section{Related work}
\label{sec:relatedWork}

Early work~\cite{dandekar2011} provides a generalized formalism for payment networks.
Closer to our subject, more recent studies~\cite{Waugh2020,vsatcs2020understanding} conduct empirical studies of the LN's reliability but do not provide a theoretical framework allowing for studying the network under different assumptions.

Multi-part payments (earlier also known as atomic multi-path payments, or AMP) have been proposed for the LN as a protocol extension~\cite{BOLTAMP} to improve reliability.
In~\cite{Piatkivskyi2018}, earlier work~\cite{dandekar2011} has been adapted and experimentally tested on a model for the LN, confirming higher success probability for smaller payment amounts.
The authors show the advantages (higher success rate and lower capital requirements) and limitations (lack of atomicity) of the proposed split payments method compared to AMP~\cite{BOLTAMP}.
Another comprehensive dynamical model~\cite{Malavolta2017} provides theoretical and simulation results based on blocking and non-blocking concurrent payment assumptions.

Multiple path-finding protocols have been proposed the the LN and payment channel networks in general, such as Flare~\cite{Prihodko2016}, SpeedyMurmurs~\cite{roos2017}, SilentWhispers~\cite{Malavolta2017a}, and Spider~\cite{Sivaraman2018spider,sivaraman2020spider}.
A model for redundant parallel payments has been proposed as Boomerang~\cite{Bagaria2019}.
Rebalancing~\cite{Pickhardt2019a} has also been suggested to improve LN reliability.

The LN has been shown to exhibit multiple privacy weaknesses due to statistical hints~\cite{Beres2019}, 
on-chain footprint~\cite{nowostawski2019eval, Tang2019}, timing measurements~\cite{Rohrer2020}, 
and other attack vectors~\cite{Kohen2019, Tang2020, Tochner2019, PerezSola2019, Mizrahi2020, Tikhomirov2020a}.
Most relevant to this work, adversarial probing of channel balances has been initially introduced in~\cite{HerreraJoancomarti2019} and unproved upon in~\cite{Dam2019,Nisslmueller2020,Kappos2020}
We use a method close to the one of~\cite{Tikhomirov2020}, supporting precise error handling for multi-hop payments.
We note that channel probing is not necessarily adversarial: nodes, in fact, probe paths in the course of normal payments.
Some LN wallets such as Zap~\cite{ZapSource} internally send probes to check route availability.

\section{Conclusion}
\label{sec:conclusion}

In this work, we have introduced and applied a mathematical framework to model the uncertainty of channel balances in the LN using probability theory.
Based on the model, we make two practical recommendations that LN nodes can follow to improve payment reliability.
First, they can now confidently decide into how many parts, if at all, it should split a payment of a given amount.
In particular, we demonstrate that splitting provides no benefit for small payments and estimate an upper bound on the achievable improvement with MPP.
Second, nodes can considerably decrease the average number of expected payment attempts by sorting paths by their prior success probabilities.
Furthermore, we use our framework to demonstrate the effectiveness of the BOLT14 proposal for a channel rebalancing protocol.
The framework can be used to evaluate other potential protocol upgrades such as Boomerang~\cite{Bagaria2019}.
On a more practical note, we stress that, instead of aiming for short path lengths, nodes should minimize the number of \textit{uncertain} hops in paths, as path success probabilities decline exponentially with that number.

\section{Acknowledgements}
\label{sec:ack}
We are grateful to Fabrice Drouin for pointing out several times and in particular during the 2nd Lightning Developer Summit in 2018 in Adelaide that work should be spent on making payments more reliable.
Similarly we thank Dr. Christian Decker for helpful discussions of preliminary results and emphasizing on the point that Lightning Network developers seem uncertain about how to split payment amounts in the case of multipart payments. 
We thank Dr. Barbara Peil for a critical review of the probabilistic model and her useful comments and feedback.

This work is partially supported by the Luxembourg National Research Fund (FNR) project FinCrypt (C17/IS/11684537).
\typeout{} 
\bibliography{mpp}
\bibliographystyle{plain}

\appendix

\subsection{Proofs}
\label{appendix:proofs}

Proof that the Entropy of a uniform distribution defined by $P(X=b)=\frac{1}{c+1}=\log(c+1)$
\begin{equation}
\begin{aligned}
H(P) & = - \sum_{b\in \Omega} P(X = b)\cdot\log(P(X = b)) \\
& = - \sum_{b\in \Omega} \frac{1}{c+1}\cdot\log(\frac{1}{c+1}) \\
& = - \frac{1}{c+1} \sum_{b\in \Omega}\log(\frac{1}{c+1}) \\
& = - \frac{1}{c+1} \cdot \log(\frac{1}{c+1}) \sum_{b\in \Omega}1 \\
& = - \frac{1}{c+1} \cdot \log(\frac{1}{c+1}) \cdot (c+1) \\
& = - \log(\frac{1}{c+1})\\
& = \log(c+1)
\end{aligned}
\end{equation}

In the case of uniform distributions, the information gain from a failed payment of size $a$ in a channel of capacity $c+1$ is just the difference in Entropy values: 
\begin{equation}
  \delta = \sum_{i=0}^{a-1}\frac{1}{a}\log\left(\frac{c+1}{a}\right) = \log(c+1) - \log(a)
\end{equation}

\subsection{Success probabilities for the mixed model}
As the probe of the full network actually revealed a mixed model for the channel balance distributions, we show how this would be used in our model.
The path success probability $s$ of a payment with size $a$ along $l$ uncertain channels of capacity $c$ in the mixed model is defined as:
\begin{equation}
  s(a,c,p) = \left(\frac{1}{2}\right)^{p\cdot l}\cdot \left(  \frac{c - a + 1}{c+1} \right)^{(1-p)\cdot l}
\end{equation}
where $p$ is the probability for a channel to follow the bimodal distribution and $(1-p)$ is the probability for the channel balance to follow the uniform distribution.
We can see in the extreme cases of $p=1$, we have the probability of every channel having the balance on the correct side, which is $\left(\frac{1}{2}\right)^l$ showing no relation to the size of the payment.
In the case where $p=0$, we do have just the probability that each channel is uniformly distributed which is $\left(  \frac{c - h + 1}{c+1} \right)^{l}$.
In the following diagram, we can see the success probabilities of payment attempts in the mixed model where the size of the HTLC was chosen to be $10\%$ of the channel capacity.
We see that the uniform distribution yields higher success rates than the bi-modal distribution, with the mixed model in the middle.
Of course, the order of the curves would be reversed in the case where the HTLC's are larger than $50\%$ of the channel capacity.

\includegraphics[width=0.5\textwidth]{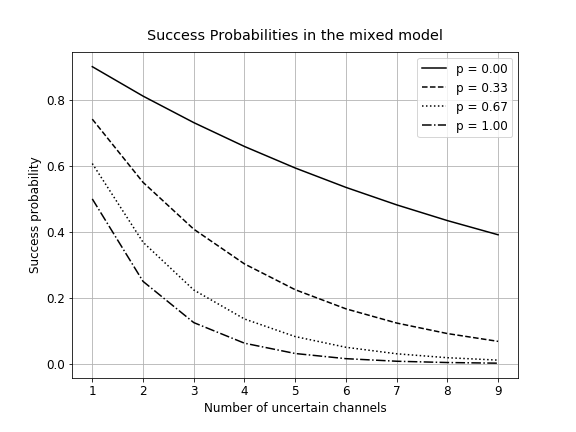}
\captionof{figure}{}


\end {document}